\documentclass[a4paper]{jpconf}
\usepackage{graphicx,amsmath,amssymb,amsbsy,amsthm,epstopdf}
\begin{document}
\title{Interpreting Quantum Discord in Quantum Metrology}

\author{Davide Girolami}

\address{Department of Atomic and Laser Physics, University of Oxford, Parks Road, Oxford OX1 3PU, UK}

\ead{davegirolami@gmail.com}

\begin{abstract}
Multipartite quantum systems show properties which do not admit a classical explanation. In particular, even nonentangled states can enjoy a kind of quantum correlations called quantum discord. I discuss some recent results on the role of quantum discord in metrology. Given an interferometric phase estimation protocol where the Hamiltonian is initially unknown to the experimentalist, the quantum discord of the probe state quantifies the minimum precision of the estimation. This provides a physical interpretation to a widely investigated information-theoretic quantity.
\end{abstract}

\section{Introduction}
\begin{figure}[h]
\centering
\includegraphics[scale=.4]{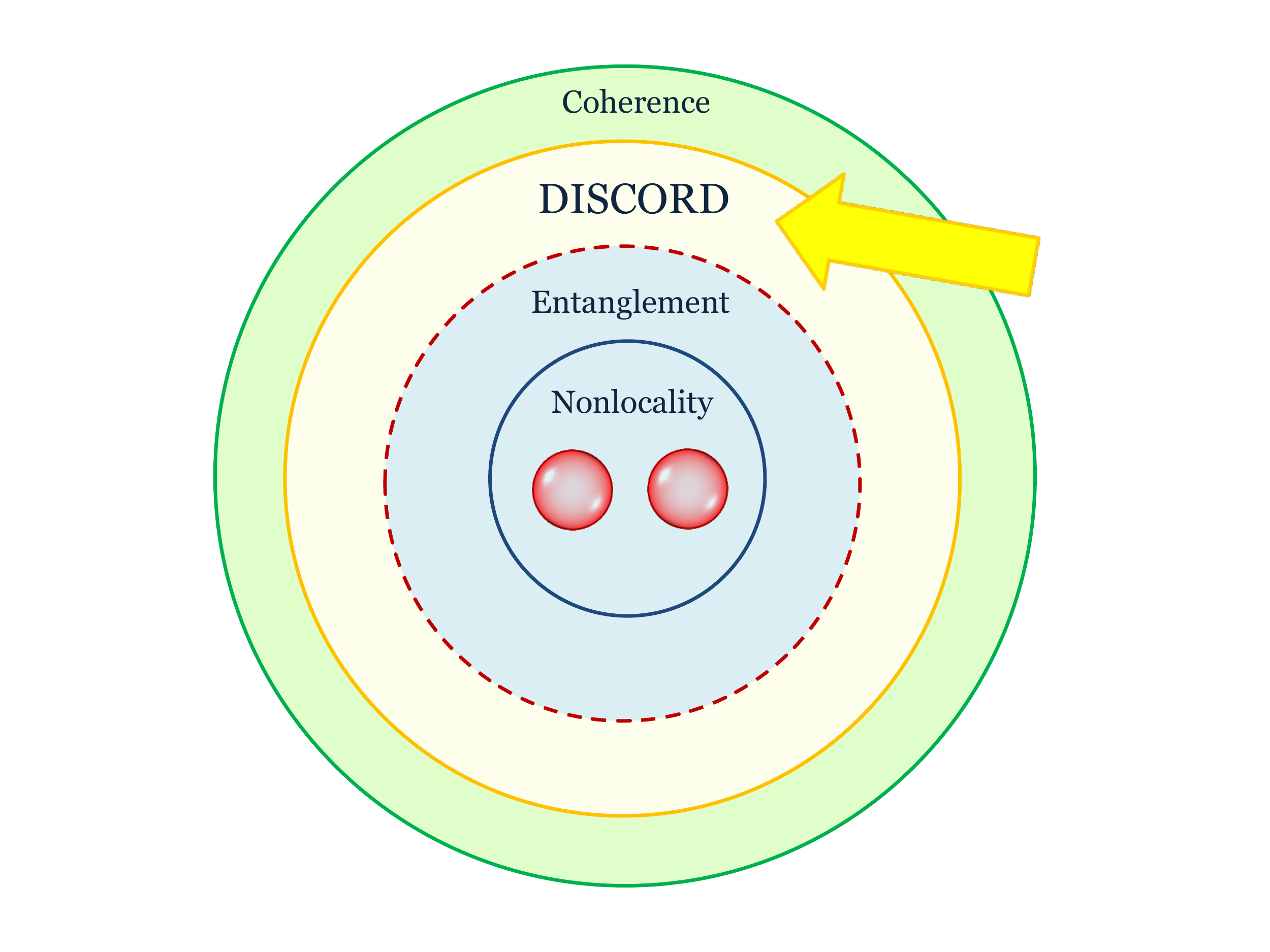}
\caption{\label{circle} Hierarchy of quantum features for bipartite mixed states, here depicted as two spheres, whilst the argument can be generalised to the multipartite case.  Coherence is the necessary ingredient for any quantum behaviour. While pure states are divided in entangled and classically correlated states, the correlation structure of mixed states is richer. Non-locality, i.e. violation of Bell inequalities, is a property enjoyed by a subset of entangled states,  and (almost all) separable states show a nonclassical statistical dependence called quantum discord (from now on, I drop the quantum label). Here I will discuss an operational interpretation of discord in quantum metrology (steering and contextuality, other spooky features of compound quantum systems, are not here depicted for the sake of simplicity).}
\end{figure}

 Developing new technology requires to encode an increasing amount of information in a decreasing amount of space, and then to process it as fast as possible.  This demand for miniaturisation and for speed makes quantum effects relevant for information manipulation, as the physical carriers of the information reach the quantum regime, e.g. atomic scales. In spite of a century of successes of the quantum mechanical predictions, characterizing the boundary between the classical and quantum worlds is still and intriguing challenge for quantum physicists. Apart from improving our ability to store, manipulate and transmit data, a better understanding of the transition between classical and quantum scenarios may convey a deeper grasp of some biological processes and more generally of complex systems.
In order to achieve these objectives, we need to identify the key particulars of a quantum phenomenon. Coherence seems to be the premier evidence of quantumness, emerging from wave-like probability distributions of measurement outcomes. Yet, subtler quantum features emerge when multipartite systems are considered. Quantum entanglement is a statistical dependence between parts of composite systems which does not admit an explanation in classical Physics. It has been intensively studied, and it is usually considered the most promising resource to provide speed-up to information protocols \cite{horo}.  Taken for granted the ability to prepare systems in pure states, which is a somehow idealised situation,  entanglement is the only kind of quantum correlations (in fact, all but a null measure set of pure states is entangled). \\
However, large scale technology needs to deal with high temperature, high disorder systems, whose description is provided by mixed states. Quantum correlations in mixed states are generally of difficult characterization, as an hierarchical structure emerges, see Fig.~\ref{circle}. Also, it is  unclear if entanglement keeps the status of  premier quantum resource for mixed state quantum computing. Reaching a final answer to this question is an extremely challenging task, as well as to identify alternative quantum resources for mixed state quantum technology. 

A more modest goal is to determine if nonentangled states still enjoy some form of quantumness.
Technically, almost all separable states of multipartite systems are inherently quantum: they show a kind of quantum correlations called discord \cite{OZ,HV}, which is not observable when the state of the system is described by a classical probability distribution.\\
For pure states, discord and entanglement coincide, while for mixed states the interplay between these two quantum features has been extensively investigated (see \cite{reviewdiscord} and references therein). 
More generally, discord has been studied on its own because of some interesting properties: it can be created by local operations and classical communications (LOCC) and it is intrinsically robust under noisy dynamics. Several operational interpretations of discord have been proposed in the recent years \cite{reviewdiscord}. I here report the original definition \cite{OZ,HV}. Given, for the sake of simplicity, a bipartite system $AB$, one makes a local measurement on one of the subsystems, say $A$. The state of the system is classically correlated if and only if  the total correlations between $A$ and $B$, measured by the mutual information, are invariant.  In the quantum case, the mutual information is nonincreasing under local operations, i.e. the measurement destroys correlations.  The discord of the state is  the amount of correlations destroyed when the most classical, i.e. the least disturbing  local measurement, is performed.\\
Other interesting arguments to explain the emergence of discord as an intrinsically quantum phenomenon have been proposed. In this work, I discuss a recently developed interpretation of discord in the context of quantum metrology,  i.e. the discipline studying how Quantum Mechanics limits or improves the precision of measurements. In Section \ref{metro}, I report two recent contributions on this line \cite{lqu,qip}. First, bipartite discord is redefined as the minimum amount of genuine quantum uncertainty on a local observable. Then, I discuss a standard metrology task, i.e. an interferometric phase estimation protocol, where building up discord in the input state guarantees nonvanishing precision for any Hamiltonian generating the phase shift, a condition not satisfied by classically correlated states. This theoretical result has been experimentally tested in a Nuclear Magnetic Resonance system \cite{qip}. Moreover, I provide a simple numerical example to clarify the role of discord in phase estimation. I draw the final remarks in Section \ref{end}. 
\section{Discord as quantum uncertainty on local observables}\label{metro}
\subsection{Local Quantum Uncertainty}
The uncertainty principle is one of the pillars of Quantum Mechanics. It states that independently of the efficiency of our measurement apparatus, the product of the uncertainties on measurement outcomes of noncommuting observables cannot be arbitrarily small. In a recent work \cite{lqu}, it has been reported that whenever a system $A$ shares discord with a system $B$, then even the measurement of any {\it single} quantity on $A$ is intrinsically uncertain. Let us consider a bipartite system described a density matrix $\rho_{AB}$. One wants to measure the local observable $K^A=K_A\otimes\mathbb{I}_B$. For technical reasons explained in \cite{lqu}, it is important here to require the observable to be bounded and nondegenerate. The information content of $\rho_{AB}$ about $K^A$ is here quantified by how much the measurement of $K^A$ on the state is uncertain. If the state is an eigenvector of the observable, the measurement outcome is certain. If it is a mixture of eigenvectors of $K^A$, the uncertainty is only due to imperfect knowledge of the state. Any departure from these situations implies a genuinely quantum contribution to the measurement uncertainty. Such quantum effect is provably measured by a nonnegative, convex function of the commutator $[\rho, K^A]$. The simplest function satisfying these two properties, to the best of my knowledge, is the Wigner-Yanase skew information ${\cal I}_{wy}(\rho, K^A)=-1/2\text{Tr}\{[\sqrt{\rho},K^A]^2\}$ \cite{wy,luo}. While it is always possible to find a global observable $K^{AB}$ commuting with the global state, for local observables with a fixed spectrum $\Lambda$ one can show that the Local Quantum Uncertainty defined as 
\begin{equation}\label{skew}
{\cal D}^A(\rho_{AB})=\min\limits_{K^A_{\Lambda}}{\cal I}_{wy}(\rho_{AB},K^A_{\Lambda})   
\end{equation}
is vanishing only for the null measure set of states taking the form $\rho^C_{AB}=\sum_{i}c_i|i\rangle\langle i|_A\otimes \rho^i_{B}, \sum_i c_i=1$, where $\{|i\rangle\}$ is an orthonormal basis \cite{lqu}. This is exactly the set of states of zero discord, also called classical-quantum states \cite{reviewdiscord}. Moreover, ${\cal D}^A(\rho_{AB})$  
 satisfies all the other properties identifying reliable discord-like measures: it is invariant under local unitary transformations; it is nonincreasing under quantum operations on the unmeasured party $B$; for pure states, it is an entanglement monotone. This is somehow surprising, as discord was originally introduced in an information-theoretic context as the minimum decrease of the mutual information ${\cal I}(\rho_{AB})$ after a local measurement identified by the Kraus operators $\{M^A_i\}$: ${\cal I}(\rho_{AB})-\max\limits_{M^A_i}{\cal I}(\rho_{M^A_i(A)B})$.   
  Conversely to entanglement, discord is in general asymmetric.  The quantity ${\cal D}^A$ has been defined by considering a local measurement on $A$. For pure states, discord reduces to entanglement, thus it is symmetric, yet in general for mixed states  one has ${\cal D}^A\neq {\cal D}^B $.  
   To sum up, Quantum Mechanics predicts ineludible uncertainty even on a single observable whenever the system of interest shares discord with another system. Classical uncertainty on a measurement outcome is brought about by the state mixedness, while discord entails a different kind of uncertainty due to the disturbance induced by a local measurement. The classical-quantum states are the only states which are not affected by at least one (local) measurement. 
 
 \subsection{Quantum Interferometric Power}
In the previous section, I showed a novel link between two apparently
unrelated quantum features as (global) correlations of states and (local) uncertainty on
observables, shedding light on the ultimate meaning of discord: it triggers local
quantum uncertainty.  
It is legit to ask if this result is of any practical usefulness for quantum metrology. This discipline consists of exploring the limits of measurement precision dictated by Quantum Mechanics. There are in fact methods to take advantage of quantumness for improving the sensitivity of measurement devices in a number of case studies. I refer the reader to a recent review \cite{metrorev}, mentioning here the progresses in noisy interferometry for optical and atomic setups, and a rather fascinating programme to improve gravitational wave detectors by quantum squeezing. \\
Many of the most important metrology tasks require to estimate the value of an unknown parameter in an interferometric configuration. This is the quantum measurement apparatus {\it par excellence}:  the effect of a phase shift applied to one of the arms, while it is not directly measurable as there is no Hermitian operator  associated to the phase parameter, is estimated by (a function of) the visibility at the output.  Here quantumness manifests in measurement statistics entailing quantum uncertainty and an interference fringe pattern at the output. Since discord guarantees quantum uncertainty on local observables (for example, the Hamiltonian $H$ which generates the phase transformation), a discordant probe state always implies nonvanising precision for the measurement of the phase parameter.   I provide here a formalisation of the above argument,  showing that discord is a resource for phase estimation. \\
The estimation protocol splits into three steps: preparation of an input probe $\rho$, which has to be sensitive to the parameter variations; encoding of the information about the parameter in the probe, which it is assumed to be a unitary evolution, i.e. a phase shift $\rho_{\theta}=U_{\theta}\rho U_{\theta}^{\dagger}, U_{\theta}=e^{i H \theta}$; measurement of an appropriate observable in the output state $\rho_{\theta}$, which allows to build up an (unbiased) estimator $\hat{\theta}$.  The aim is to minimise the variance of the estimator $\hat{\theta}$. The fundamental limit imposed by Quantum Mechanics for $\nu$ runs of the protocol is given by the quantum Cram\'er-Rao bound:
\begin{eqnarray}
 \text{Var}_{\rho}(\hat{\theta}) \geq 1/[\nu {\cal F}(\rho, H)],
 \end{eqnarray}
where $\nu \gg 1$ is the number of repetitions of the experiment, and ${\cal F}$ is the quantum Fisher Information (QFI) \cite{helstrom,holevo}, which evaluates the sensitivity of the probe state to the phase shift assuming that the best measurement strategy is performed.   For single parameter estimation,  the best estimator statistical spread saturates (in the asymptotic limit) the quantum Cram\'er-Rao bound:  $\text{Var}_{\rho}(\hat{\theta}_{\rm best}) = 1/[\nu {\cal F}(\rho, H)]. $

Let me now suppose that the estimation is blind, in the sense that the experimenter, call it Alice,  must prepare the input state $\rho_A$ without any prior information about the Hamiltonian. I assume that the phase direction is unveiled to her at the output, thus Alice is still allowed to carry out the most informative measurement and build the best possible estimator.  As the sensitivity of the probe is given by the amount of coherence with respect to the eigenbasis of the Hamiltonian, there is no input which  guarantees any degree of precision in the estimation. However, the situation is different if Alice  collaborates with a second player Bob and implements a two-arm interferometer to perform the estimation. While a classically correlated two-party probe $\rho_{AB}$ is still insufficient to ensure the success of the estimation regardless of the phase direction, it is provable that implementing an input state with discord always provides a fixed amount of precision. Indeed, the minimum of the QFI over all the possible same spectrum local Hamiltonians, i.e.
\begin{equation}\label{qip}
{\cal P}^A(\rho_{AB})=1/4\min\limits_{H^A_{\Lambda}}{\cal F}(\rho_{AB},H^A_{\Lambda}),   
\end{equation}
called Quantum Interferometric Power (QIP), is a {\it bona fide} measure of discord-like correlations ($1/4$ is just a normalisation factor). This is not surprising, as the QFI and the Wigner-Yanase information are parent quantities, being two quantum versions of the classical Fisher Information and satisfying the relation ${\cal I}_{wy}\leq {\cal F}\leq 2{\cal I}_{wy}$ \cite{luo}. The Wigner-Yanase information is a brute force quantisation of the classical Fisher Information definition, where a density matrix replaces a probability distribution \cite{luo}. The QFI is obtained by a further maximisation of the information retrieved by the measurement at the output of the estimation protocol \cite{helstrom,holevo}.

\subsection{A numerical example}
Here I provide a simple numerical example, which is a slightly modified version of the one reported in \cite{qip}. I compare the performance in a single run phase estimation protocol of two-qubit probe states which show a different kind of correlations between their parts.

The first one is a one-parameter Bell-diagonal state
\begin{eqnarray}
\rho_{Q}= \left(
\begin{array}{cccc}
 \frac{p+1}{4} & 0 & 0 & \frac{1}{4} p (p+1) \\
 0 & \frac{1-p}{4} & -\frac{1}{4} (p-1) p & 0 \\
 0 & -\frac{1}{4} (p-1) p & \frac{1-p}{4} & 0 \\
 \frac{1}{4} p (p+1) & 0 & 0 & \frac{p+1}{4} \\
\end{array}
\right),\nonumber
\end{eqnarray}
which  is prepared by initialising two single qubit states in $\rho_{i}=1/2 (\mathbb{I}_2+ p \sigma_z), i=A,B, p\in [0,1]$, then applying a Hadamard gate on the first qubit followed by a two-qubit CNOT gate \cite{nielsen}.

\begin{figure}[h!]
\centering
\includegraphics[scale=.8]{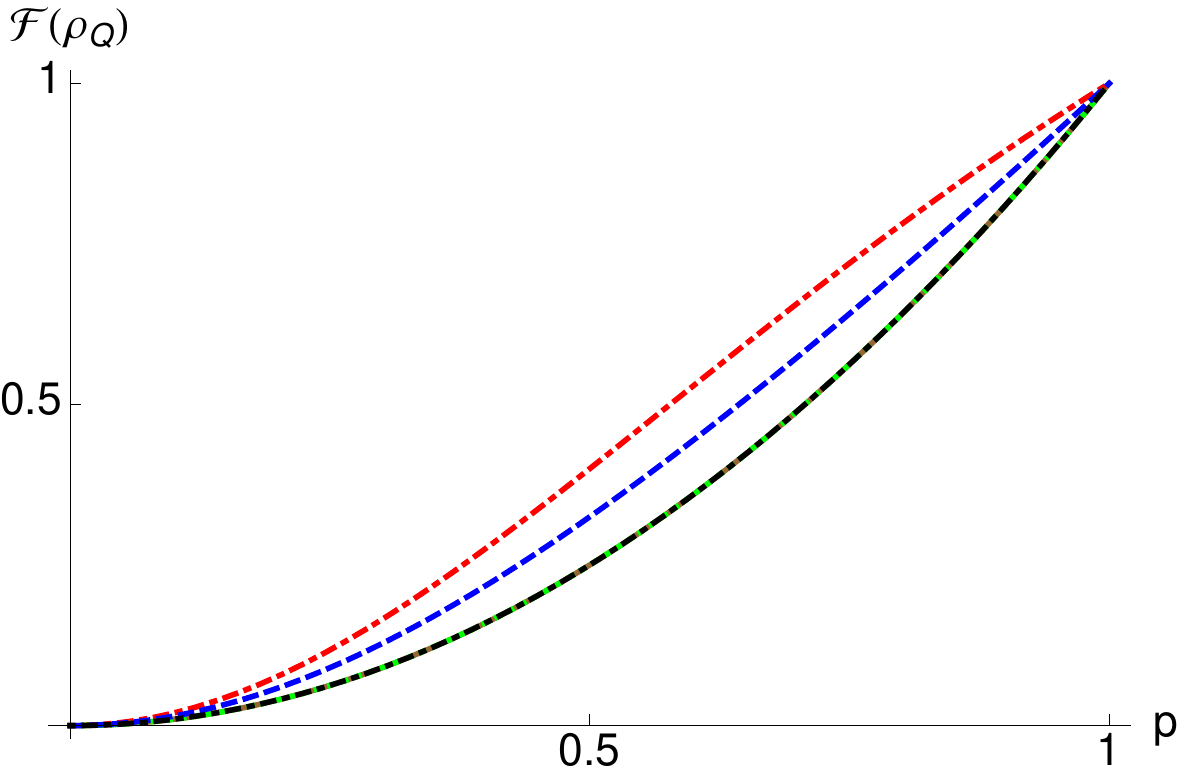}
\caption{\label{plot2}(Colors online) Quantum Fisher Information of the quantum correlated state $\rho_Q$ with respect to four local Hamiltonians $H^A=H_A\otimes\mathbb{I}_B, H^A=\sigma^A_{x,y,z},(\sigma^A_x+\sigma^A_y)/\sqrt{2}$. The parameter $p$ determines the purity of the state. I depict the functions ${\cal F}(\rho_Q,\sigma_x^A)=p^2$ (brown, continuous line), ${\cal F}(\rho_Q,\sigma_y^A)=2 p^2/(1 + p^2)$ (red dot-dashed line), ${\cal F}(\rho_Q,\sigma_z^A)=p^2$ (green dotted line), ${\cal F}(\rho_Q,(\sigma^A_x+\sigma^A_y)/\sqrt{2})=p^2 (p^2+3)/(2 (p^2+1))$ (blue dashed line), and the QIP given by the formula in \cite{qip} (black dot-dashed line). Note that the QFI is lower bounded by the discord of the state, which therefore ensures a minimum precision to the estimation.}
\end{figure}
\begin{figure}[h!]
\centering
\includegraphics[scale=.9]{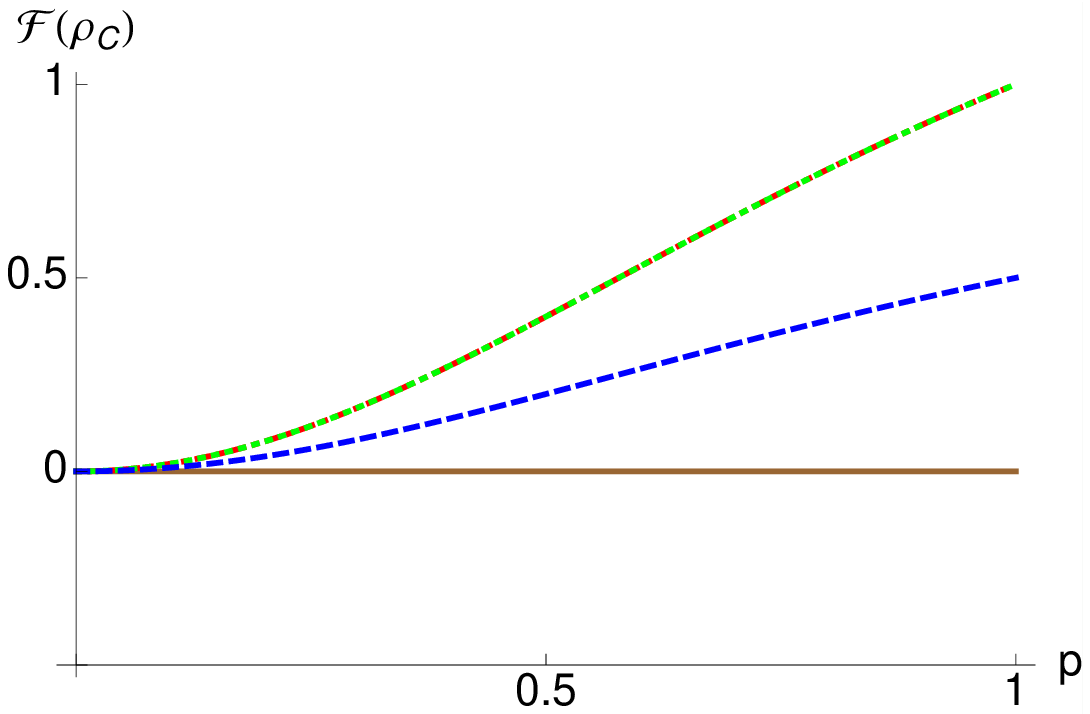}
\caption{\label{plot1}(Colors online) Quantum Fisher Information of the classically correlated state $\rho_C$ with respect to four local Hamiltonians $H^A=H_A\otimes\mathbb{I}_B, H^A=\sigma^A_{x,y,z},(\sigma^A_x+\sigma^A_y)/\sqrt{2}$. The parameter $p$ determines the purity of the state. I depict the functions ${\cal F}(\rho_C,\sigma_x^A)=0$ (brown, continuous line), ${\cal F}(\rho_C,\sigma_y^A)=2 p^2/(1 + p^2)$ (red dot-dashed line), ${\cal F}(\rho_C,\sigma_z^A)=2 p^2/(1 + p^2)$ (green dotted line), ${\cal F}(\rho_C,(\sigma^A_x+\sigma^A_y)/\sqrt{2})=p^2/(1 + p^2)$ (blue dashed line). Note that the state is unable to store information about the phase shift along the $x$-direction.}
\end{figure}

The second probe is a classically correlated state:
\begin{eqnarray}
\rho_C=\left(
\begin{array}{cccc}
\frac 14& \frac{p^2}{4}& \frac p4&  \frac p4\\
 \frac{p^2}{4}& \frac 14&\frac p4 & \frac p4 \\
\frac p4& \frac p4& \frac 14& \frac{p^2}{4}\\
\frac p4&\frac p4&\frac{p^2}{4}&\frac 14
\end{array}
\right). 
\end{eqnarray}

Note that the comparison is fair as the probes have the same purity: $\text{Tr}\{\rho_{C,Q}^2\}=1/4 (1+p^2)^2$. 
I calculate the QFI for the two states with respect to four local observables $H_A\otimes \mathbb{I}_B, H_A=\sigma_{x,y,z},(\sigma_{x}+\sigma_{y})/{\sqrt 2}$ . This is given by  ${\cal F}(\rho_{Q,C}, H^A)= 4 \sum_{p_i+p_j\neq0} \frac{(p_i-p_j)^2}{p_i + p_j} |\langle\psi_i|H^A|\psi_j\rangle|^2$, where $\{p_i\},\{\psi_i\rangle\}$ are the eigenvalues and eigenvectors of $\rho_{Q,C}$. The QIP for the quantum correlated state is obtained by applying the closed formula derived in the appendix of \cite{qip}, while of course one has ${\cal P}^A(\rho_{C})=0$. A study of how the QFI and the QIP change with variations of the noise parameter $p$ is presented in Figs.~\ref{plot2},\ref{plot1}. The results highlight the peculiarity of the discordant state: it guarantees a nonvanishing precision in the estimation for any phase direction.

 \section{Conclusions}\label{end}
I here discussed recent advances on the interpretation of discord in metrology. The peculiar asymmetry of this kind of quantum correlations between parts of a compound system is explained by means of an interferometric phase estimation scheme.  Whenever a system shares discord, Quantum Mechanics predicts that any local measurement has a degree of uncertainty which translates into an improved sensitivity in parameter estimation. The Local Quantum Uncertainty and the Quantum Interferometric Power are parent discord-like measures which quantify the minimum amount of precision in interferometric phase estimation. An interesting question is to establish if  the metrologic measures of discord, which have been introduced to catch bipartite statistical dependence, can be extended to quantify multipartite correlations.  We want to find out when and how physical systems undergo genuinely quantum processes, and when such effects may boost the performance of information processing protocols. The discussed result may lead to new ways of exploiting quantum physics for delivering entanglement-free quantum technology and provide conceptual advances in the understanding of complex systems.

\ack
The work was supported by the UK Engineering and Physical Sciences Research Council (EPSRC) under the Grant No. EP/L01405X/1, the Foundational Questions Institute (FQXi) and the Wolfson College, University of Oxford. I thank the organisers of the Conference {\it DICE 2014 -- Spacetime - Matter - Quantum Mechanics} held in Castiglioncello for the kind hospitality.

\end{document}